\documentclass[a4paper,UKenglish,cleveref, autoref, thm-restate]{lipics-v2021}

\pdfoutput=1 
\hideLIPIcs  

\title{Combining Predicted and Live Traffic with Time-Dependent A* Potentials}

\author{Nils Werner}{Karlsruhe Institute of Technology, Germany}{}{}{}
\author{Tim Zeitz}{Karlsruhe Institute of Technology, Germany}{tim.zeitz@kit.edu}{https://orcid.org/0000-0003-4746-3582}{}

\authorrunning{N. Werner and T. Zeitz} 

\Copyright{Nils Werner and Tim Zeitz} 



\relatedversiondetails[cite=wz-cplttdap-22]{Conference Paper}{https://doi.org/10.4230/LIPIcs.ESA.2022.89} 

\supplementdetails[subcategory={Source Code},swhid={swh:1:dir:f0d810343d53d5ad87db428e89a32cb038dcaeab;
origin=https://github.com/kit-algo/tdpot;
visit=swh:1:snp:dbc8029b1b440b545865e95431d48f93fdc3afd1;
anchor=swh:1:rev:fbbe618f87a604a4f0c929276203f07b6a5b1a94}]{Software}{https://github.com/kit-algo/tdpot}

\acknowledgements{
We want to thank Jonas Sauer for many helpful discussions on algorithmic ideas and proofreading of drafts of this paper.
Further, we also want to thank the anonymous reviewers for their helpful comments.
}

\nolinenumbers 

\EventEditors{Shiri Chechik, Gonzalo Navarro, Eva Rotenberg, and Grzegorz Herman}
\EventNoEds{4}
\EventLongTitle{30th Annual European Symposium on Algorithms (ESA 2022)}
\EventShortTitle{ESA 2022}
\EventAcronym{ESA}
\EventYear{2022}
\EventDate{September 5--9, 2022}
\EventLocation{Berlin/Potsdam, Germany}
\EventLogo{}
\SeriesVolume{244}
\ArticleNo{9}

\newcommand*{\pred}{p}
\newcommand*{\comb}{c}

\newcommand*{\dist}{\mathcal{D}}

\newcommand*{\live}{\ell}

\newcommand*{\tdep}{\tau^{\operatorname{dep}}}
\newcommand*{\tend}{\tau^{\operatorname{end}}}
\newcommand*{\tnow}{\tau^{\operatorname{now}}}
\newcommand*{\tmax}{\tau^{\max}}

\newcommand*{\gchu}{G^{\uparrow}}
\newcommand*{\gchd}{\overleftarrow{G^{\downarrow}}}
\newcommand*{\echu}{E^{\uparrow}}
\newcommand*{\rechd}{\overleftarrow{E^{\downarrow}}}
\newcommand*{\echd}{E^{\downarrow}}

\newcommand*{\pcfn}{\underline{b'^+}}
\newcommand*{\bucketlen}{\beta}

\usepackage[algo2e,vlined]{algorithm2e}
\usepackage{booktabs}

\usepackage{tikz}
\tikzstyle{node}=[circle,inner sep=0.5mm,minimum size=6.5mm,draw = black]
\tikzstyle{snode}=[circle,minimum size=2mm,fill = black]
\usetikzlibrary{shapes.misc}
\tikzset{cross/.style={cross out, draw=black, minimum size=2*(#1-\pgflinewidth), inner sep=0pt, outer sep=0pt},
cross/.default={1pt}}

\begin{document}

\maketitle

\begin{abstract}
We study efficient and exact shortest path algorithms for routing on road networks with realistic traffic data.
For navigation applications, both current (i.e., live) traffic events and predictions of future traffic flows play an important role in routing.
While preprocessing-based speedup techniques have been employed successfully to both settings individually, a combined model poses significant challenges.
Supporting predicted traffic typically requires expensive preprocessing while live traffic requires fast updates for regular adjustments.
We propose an A*-based solution to this problem.
By generalizing A* potentials to time dependency, i.e.\ the estimate of the distance from a vertex to the target also depends on the time of day when the vertex is visited, we achieve significantly faster query times than previously possible.
Our evaluation shows that our approach enables interactive query times on continental-sized road networks while allowing live traffic updates within a fraction of a minute.
We achieve a speedup of at least two orders of magnitude over Dijkstra's algorithm and up to one order of magnitude over state-of-the-art time-independent A* potentials.
\end{abstract}

\section{Introduction}

An important feature of modern routing applications and navigation devices is the integration of traffic information into routing decisions.
The more comprehensive the considered traffic information, the better the suggested routes, the more accurate the predicted arrival times and ultimately, the more satisfied the users.
For routing, we can distinguish between two aspects of traffic:
On the one hand, there are \emph{predictable} traffic flows.
For example, certain highways will consistently have traffic jams on weekday afternoons due to commuters driving home.
On the other hand, unexpected events such as accidents may also have significant influence on the \emph{current} (i.e., \emph{live}) traffic situation.
While it may be sufficient to focus on the current traffic situation to answer short-range routing requests, mid- and long-range queries require taking both types of traffic into account.
We therefore aim to provide routing algorithms which incorporate \emph{combined} predicted and live traffic information.

A common approach for routing in road networks is to model the network as a directed graph where intersections are represented by vertices and road segments by edges.
With edge weights representing travel times, routing requests can be answered by solving the classical shortest path problem.
When considering predicted traffic, edge weights can be modeled as functions of the time of day, which is commonly referred to as \emph{time-dependent routing}.
Dijkstra's algorithm can be used to solve these problems exactly and, at least from a theoretical perspective, efficiently~\cite{d-ntpcg-59}.
However, on the continental-sized road networks used in modern routing applications, it may take seconds to answer mid- or long-range queries, which is too slow for most practical applications.
We therefore study algorithms to compute shortest paths significantly faster than Dijkstra's algorithm while retaining exactness.

One approach to accelerate Dijkstra's algorithm is goal-directed search, i.e.\ the A* algorithm~\cite{hnr-afbhd-68}.
Dijkstra's algorithm uses a priority queue to explore vertices by ascending distance from the source until it reaches the target.
A* changes this slightly and employs a \emph{potential} function which estimates the remaining distance from vertices to the target to change the queue order and explore vertices closer to the target earlier.
The performance of A* crucially depends on the tightness of these estimates.
The core algorithmic idea of this work is to use A* with time-dependent potential functions, i.e.\ we obtain tighter estimates and therefore faster queries by taking the time of day when a vertex is visited into account.

\subsection{Related Work}

Efficient and exact route planning in road networks has received a significant amount of research effort in the past decade.
Since a comprehensive discussion is beyond the scope of this paper, we refer to~\cite{bdgmpsww-rptn-16} for an overview.
An approach that has proven effective is to exploit the fact that usually many queries have to be answered on the same network, which rarely changes.
Thus, these queries can be accelerated by computing auxiliary data in an off-line preprocessing phase.

A popular technique which follows this approach is \emph{Contraction Hierarchies} (CH)~\cite{gssv-erlrn-12}.
During the preprocessing vertices are ranked heuristically by ``importance'' where more important vertices are part of more shortest paths.
Shortcut edges are inserted to skip over unimportant vertices.
This allows for a very fast query where only a few vertices are explored.
On typical continental-sized networks, queries take well below a millisecond.
\emph{Multi-Level Dijkstra} (MLD)~\cite{swz-umlgt-02} is a similar approach that also utilizes shortcut edges but inserts them based on a multi-level partitioning.
It achieves slightly slower query times of around a millisecond.
MLD is the first algorithm to operate under the \emph{Customizable Route Planning} (CRP) framework~\cite{dgpw-crprn-13}, i.e.\ it has a second, faster preprocessing phase called \emph{customization} which allows integrating arbitrary weight functions (or live traffic updates for the current weights) into the auxiliary data without rerunning the entire first preprocessing phase, which is much slower.
The MLD customization takes a few seconds, which allows for running it every minute.
This three-phase setup has proven to be instrumental to support live traffic scenarios in practical applications~\cite{bingblog}.
Therefore, CH was generalized to Customizable Contraction Hierarchies (CCH)~\cite{dsw-cch-15} to support customizability as well.

Both CH and MLD have been extended to time-dependent routing.
However, dealing with weight functions instead of scalar weights makes the preprocessing much harder and leads to difficult trade-offs.
TCH~\cite{bgsv-mtdtt-13} has fast queries but a very expensive preprocessing phase (up to several hours) and may produce prohibitive amounts of auxiliary data ($> 100$\,GB).
TD-CRP~\cite{bdpw-dtdrp-16}, an extension of MLD, even follows a three-phase approach and has a relatively fast customization phase.
However, this is only possible by giving up exactness.
Also, TD-CRP does not support path unpacking.
CATCHUp~\cite{swz-sfert-21} adapts CCH to the time-dependent setting and has fast and exact queries with significantly reduced memory consumption.
While it has a customization phase, running it takes significantly longer than a traditional CCH or CRP customization.
On the networks used in this paper, a CATCHUp customization may even take hours, which is too slow for a setting with live traffic updates.
Time-dependent Sampling (TDS)~\cite{strasser:OASIcs:2017:7897} is another CH-based approach.
While TDS does support both predicted and dynamic traffic information, it cannot guarantee exactness.

ALT~\cite{gh-cspas-05,gw-cppsp-05} is an early A*-based speedup technique for routing in road networks.
It combines precomputed distances to a few \emph{landmark} vertices with the triangle inequality to obtain distance estimates to the target vertex.
However, query times are significantly slower than with shortcut-based approaches such as CH or MLD.
ALT also has been extended to dynamic and time-dependent settings~\cite{dn-crdtd-12}. 
While this approach allows incremental modifications of the input travel times, it is not as flexible as customization based approaches allowing arbitrary updates.

\emph{CH-Potentials}~\cite{strasser_et_al:LIPIcs.SEA.2021.6} is another more recent A*-based approach.
CH-Potentials use Lazy RPHAST~\cite{strasser_et_al:LIPIcs.SEA.2021.6}, an incremental many-to-one CH query variant, to compute exact distances toward the target.
This allows for tighter estimates and faster queries than what is possible with ALT.
CH-Potentials can be applied to a variety of routing problem variants.
The original publication even mentions a combination of live and predicted traffic.
However, the reported query times are above 100\,ms.
We consider this too slow for practical applications.

\subsection{Contribution}

In this work, we introduce a time-dependent generalization of A* potentials.
We present two Lazy RPHAST extensions that realize a time-dependent potential function and discuss how to apply them to queries in a setting that combines live and predicted traffic.
An extensive evaluation confirms the effectiveness of our potentials.
Queries incorporating both predicted and current traffic can be answered within few tens of milliseconds.
Live traffic updates can be integrated within a fraction of a minute.
Our time-dependent potentials are up to an order of magnitude faster than CH-Potentials and about two orders of magnitude faster than Dijkstra's algorithm.
To the best of our knowledge, this makes our approach the first to achieve interactive query performance while allowing fast updates in this setting.

\section{Preliminaries}
We consider simple directed graphs $G=(V,E)$ with $n=|V|$ vertices and $m=|E|$ edges.
We use $uv$ as a short notation for an edge from a \emph{tail} vertex $u$ to a \emph{head} vertex $v$.
Weight functions $w : E \to (\mathbb{Z} \to \mathbb{N}^0)$ map edges to time-dependent functions, which in turn map a departure time $\tau$ at the tail $u$ to a travel time $w(uv)(\tau)$.
To simplify notation, we will often write $w(uv, \tau)$.
When it is clear from the context that we are writing about constant functions, we omit the time argument and write $w(uv)$.
The reversed graph $\overleftarrow{G} = (V, \overleftarrow{E})$ contains a reversed edge $vu$ for every edge $uv \in E$.
In this paper, we only need time-independent corresponding reversed weight functions.
Therefore, we define $\overleftarrow{w}(vu) = w(uv)$.
The travel time of a path $P = (v_1,\dots,v_k)$ is defined recursively $w(P, \tau) = w(v_1 v_2, \tau) + w((v_2,\dots,v_k), \tau + w(v_1 v_2, \tau))$ with the base case of an empty path having a travel time of zero.
A path's travel time can be obtained by successively evaluating travel times of the edges of the path.
We denote the travel time of a \emph{shortest} path between vertices $s$ and $t$ for the departure time $\tdep$ as $\dist_w(s,t,\tdep)$.
We assume that all travel time functions adhere to the \emph{First-In First-Out} (FIFO) property, i.e.\ departing later may never lead to an earlier arrival.
Formally stated, this means $\tau + w(\tau) \leq \tau + \epsilon + w(\tau + \epsilon)$ for any $\epsilon \geq 0$.
With non-FIFO travel time functions, the shortest path problem becomes \textsf{NP}-hard~\cite{or-tnp-89,z-nphsp-22}.

\subsection{Problem Model}

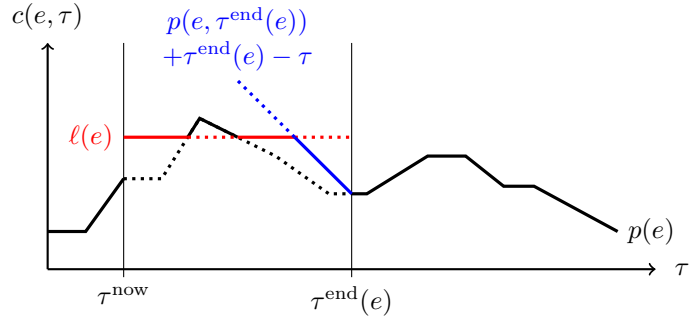
\begin{figure}
\centering
\begin{tikzpicture}[]
\draw[->, thick] (0,1) -- (8,1);
\draw[->, thick] (0,1) -- (0,4);

\draw[] (1,4) -- (1,1cm-3pt) node[anchor=north] {$\tnow$};
\draw[] (4,4) -- (4,1cm-3pt) node[anchor=north] {$\tend(e)$};

\node[] () at (8.35, 1) { $\tau$ };
\node[] () at (0, 4.35) { $\comb(e, \tau)$ };

\draw[very thick] (0,1.5) -- (0.5,1.5) -- (1, 2.2);
\draw[very thick,dotted] (1,2.2) -- (1.5,2.2) -- (2,3) -- (3,2.5) -- (3.7, 2) -- (4,2);
\draw[very thick] (1.84375,2.75) -- (2,3) -- (2.5,2.75);
\draw[very thick] (4,2) -- (4.2,2) -- (5,2.5) -- (5.5,2.5) -- (6,2.1) -- (6.4,2.1) -- (7.5,1.5) node[anchor=west] {$\pred(e)$};

\draw[very thick,red] (1.84375,2.75) -- (1,2.75) node[anchor=east,red] {$\live(e)$};
\draw[very thick,red,dotted] (1.84375,2.75) -- (2.5,2.75);
\draw[very thick,red] (2.5,2.75) -- (3.25,2.75);
\draw[very thick,red,dotted] (3.25,2.75) -- (4,2.75);

\draw[very thick,dotted,blue] (3.25,2.75) -- (2.5,3.5) node[anchor=south,blue, align=center] {$\pred(e, \tend(e))$\\$+ \tend(e) - \tau$};
\draw[very thick,blue] (3.25,2.75) -- (4,2);
\end{tikzpicture}
\caption{
Combined travel time function $\comb(e, \tau) = \max(\pred(e, \tau), \min(\live(e), \pred(e, \tend(e)) + \tend(e) - \tau))$ with both predicted and live traffic information.
The predicted traffic $\pred(e)$ is indicated in black.
The live travel time $\live(e)$ with expected end $\tend(e)$ is depicted in red.
The switch back to the predicted function is colored in blue.
The solid line indicates the combined function $\comb(e)$ for the current day.
For later days, only $\pred$ will be used.
Dotted lines only serve the purpose of visualization.
}
\label{fig:combined_tt}
\end{figure}

We consider an application model with three phases.
During the \emph{preprocessing} phase, the graph $G=(V,E)$ and a weight function $\pred$ of time-dependent traffic predictions are given.
Predicted travel time functions are periodic piecewise linear functions represented by a sequence of breakpoints covering one day.
A preprocessing algorithm may now precompute auxiliary data, which may take several hours.
In the \emph{update} phase, a weight function $\live$ of currently observed live travel times are given for the current moment $\tnow$.
These live travel times are time-independent and can be represented by a single scalar value.
Further, each edge $e$ has a point in time $\tend(e)$ when we switch back to the predicted travel time.
For edges without live traffic data, we set $\tend(e) = \tnow$.
We assume that traffic predictions are conservative estimates and that live traffic will only be slower than the predicted traffic due to accidents and other traffic incidents, i.e.\ $\pred(e, \tnow) < \live(e)$.
Therefore, we define the combined travel time function $\comb(e, \tau) = \max(\pred(e, \tau), \min(\live(e), \pred(e, \tend(e)) + \tend(e) - \tau))$.
It follows that $\pred(e, \tau) \leq \comb(e, \tau)$.
The update phase will be repeated frequently and should therefore be as fast as possible.
During the final \emph{query} phase, many shortest path queries $(s,t,\tdep)$ where $\tdep \geq \tnow$ should be answered as quickly as possible by obtaining a path $P = (s,\dots,t)$ that minimizes $\comb(P, \tdep)$.
Figure~\ref{fig:combined_tt} depicts an example of such a combined travel time function.

Our model has two important restrictions.
First, the dynamic real-time traffic information $\live$ is handled separately from the traffic predictions $\pred$.
Fast updates to the predicted traffic functions $\pred$ are not the goal of our work.
While this might seem less flexible than dynamic traffic predictions $\pred$, we believe that our model is actually more practical.
This is because the traffic predictions are \emph{periodic} functions.
But live traffic incidents are inherently tied to the current moment and are not expected to repeat in 24 hours.
Second, our model assumes predicted traffic to be a lower bound of the real-time traffic.
If the observed live travel time were faster than the predicted travel time, the live travel time would be ignored.
While this a severe restriction from a theoretical perspective, it is only a minor limitation for our practical problem.
Live traffic should account for unexpected traffic events which will almost always only make traffic worse.
If the live traffic is frequently better than the predicted traffic, the predictions should be adjusted at some point.
We discuss these restrictions further and compare our problem to similar models in related work in Appendix~\ref{sec:appendix:model}.

\subsection{Fundamental Algorithms}

\emph{Dijkstra's algorithm}~\cite{d-ntpcg-59} computes $\dist_w(s,t,\tdep)$ by exploring vertices in increasing order of distance from $s$ until $t$ is reached.
The distances from $s$ to each vertex $u$ are tracked in an array $\mathtt{D}[u]$, initially set to $\infty$ for all vertices.
A priority queue of vertices ordered by their distance from $s$ is maintained.
The priority queue is initialized with $s$ and $\mathtt{D}[s]$ set to $\tdep$.
In each iteration, the next closest vertex $u$ is extracted from the queue and \emph{settled}.
Outgoing edges $uv$ are \emph{relaxed}, i.e.\ the algorithm checks if $\mathtt{D}[u] + w(uv, \mathtt{D}[u])$ improves $\mathtt{D}[v]$.
If so, the queue position of $v$ is adjusted accordingly.
Once $t$ has been settled, the final distance is known, and the search terminates.
We denote visited vertices as the \emph{search space} of a query.

A*~\cite{hnr-afbhd-68} is a goal-directed extension of Dijkstra's algorithm.
It applies a \emph{potential} function $\pi_t$ which maps vertices to an estimate of the remaining distance to $t$.
This estimate is added to the queue key.
Thus, vertices closer to the target are visited earlier, and the search space becomes smaller.
It can be guaranteed that A* has computed the shortest distance once $t$ is settled when the estimates of the potential function are lower bounds of the remaining distances.
However, with only lower bound potentials, the theoretical worst case running time of A* is exponential.
Therefore, a stronger correctness property is often used.
A potential function is called \emph{feasible} if $w(uv) - \pi_t(u) + \pi_t(v) \geq 0$ for any edge $uv \in E$.
Feasibility guarantees correctness and polynomial running time.
When the potential of the target is zero, i.e.\ $\pi_t(t) = 0$, it also implies the lower bound property.

\emph{Contraction Hierarchies} (CH)~\cite{gssv-erlrn-12} is a speedup technique to accelerate shortest path searches on time-independent road networks through precomputation.
During the preprocessing, a total order $v_1 \prec \dots \prec v_n$ of all vertices $v_i \in V$ by ``importance'' is determined heuristically, where more important vertices should lie on more shortest paths.
Then, an \emph{augmented graph} $G^+ = (V, E^+)$ with additional \emph{shortcut edges} and weights $w^+$ is constructed.
Shortcut edges $uv$ allow to ``skip over'' paths $(u,\dots,x_i,\dots,v)$ where $x_i \prec u$ and $x_i \prec v$.
Therefore, $w^+(uv)$ is assigned the length of the shortest such path.
We sometimes split $G^+$ into an upward graph $\gchu = (V, \echu)$ which contains only edges $uv$ where $v \succ u$ and a downward graph $G^{\downarrow} = (V, E^{\downarrow})$ defined analogously.
The augmented graph has the property that between any two vertices $s$ and $t$, there exists an \emph{up-down-path} $P$ with $w^+(P) = \dist_w(s,t)$ which uses first only edges from $\echu$ and then only edges from $\echd$.
Such a path can be found by running the bidirectional variant of Dijkstra's algorithm from $s$ on $\gchu$ and from $t$ on $\gchd$.
Because only a few vertices are reachable in this \emph{CH search space}, queries are very fast, i.e.\ about a tenth of a millisecond on continental-sized networks.

In this work we build on \emph{Customizable Contraction Hierarchies} (CCH)~\cite{dsw-cch-15}.
For CCH, the construction of the augmented graph is split into two phases.
In the first phase, the topology of the augmented graph is constructed without considering any weight functions.
It is therefore valid for \emph{all} weight functions.
In the second \emph{customization} phase, the weights $w^+$ of the augmented graph are computed for a given weight function $w$.
The customization can be parallelized efficiently~\cite{bsw-rttau-19} and takes a couple of seconds on typical networks.

\begin{algorithm2e}
\KwData{$\mathtt{D}^{\downarrow}[u]$: tentative distance from $u$ to $t$ computed by Dijkstra's algorithm on $\gchd$}
\KwData{$\mathtt{D}[u]$: memoized final distance from $u$ to $t$, initially $\bot$}
\SetKwFunction{Dist}{ComputeAndMemoizeDist}
\SetKwProg{Fn}{Function}{:}{}
\Fn{\Dist{$u$}}{
    \If{$\mathtt{D}[u] = \bot$}{
        $\mathtt{D}[u]\leftarrow \mathtt{D}^{\downarrow}[u]$\;
        \For{all edges $uv \in \echu$}{
            $\mathtt{D}[u] \leftarrow \min(\mathtt{D}[u], \Dist(v) + w^+(uv))$\;
        }
    }
    \Return{$\mathtt{D}[u]$}\;
}
\caption{Computing the distance from a single vertex $u$ to $t$ with Lazy RPHAST.}
\label{algo:lazy_rphast}
\end{algorithm2e}

\emph{Lazy RPHAST}~\cite{strasser_et_al:LIPIcs.SEA.2021.6} is a CH query variant to incrementally compute distances from many sources toward a common target.
The first step is to run Dijkstra's algorithm on $\gchd$ from $t$, similarly to a regular CH query.
The second Dijkstra search is replaced with a recursive depth-first search (DFS) which memoizes distances.
Algorithm~\ref{algo:lazy_rphast} depicts this routine which will be called for all sources.
If the distance of a vertex $u$ was previously computed, the routine terminates immediately and returns the memoized value $\mathtt{D}[u]$.
Otherwise, the distance for all upward neighbors $v$ is obtained recursively.
The final distance is the minimum over the path distances $w^+(uv) + \mathtt{D}[v]$ via the upward neighbors $v$ and the distance possibly found in the backward search $\mathtt{D}^{\downarrow}[u]$.
Using a DFS to compute shortest distances works because $\gchu$ is a directed acyclic graph.
Using the distance to $t$ obtained by Lazy RPHAST as an A* potential is called \emph{CH-Potentials}.
Just like a regular CH query, Lazy RPHAST can be used on CCH without modifications.
In~\cite{strasser_et_al:LIPIcs.SEA.2021.6} additional optimizations for A* are discussed which we also utilize.
The goal is to reduce the impact of the potential evaluation overhead by avoiding unnecessary potential evaluations, for example for chains of degree-two vertices.

\section{Time-Dependent A* Potentials}

We now propose a time-dependent generalization $\pi_t : V \to (T \to \mathbb{Z}^{\geq 0})$ of A* potentials, i.e.\ estimates are a function of the time.
This allows us to obtain tighter estimates and enables faster queries.
Analogue to classical potentials, there are properties of time-dependent potentials to consider for the correctness of A*:
\begin{itemize}
  \item Strong First-In First-Out (FIFO): $\pi_t(v, \tau) < \pi_t(v, \tau + \epsilon) + \epsilon$ for $v \in V$, $\tau > \dist_w(s,u,\tdep)$ and $\epsilon > 0$.
        This ensures that queue keys increase monotonically with the distance from $s$.
        This property has no time-independent equivalent because it holds trivially in this case.
  \item Feasibility: $w(uv, \tau) + \pi_t(v, \tau + w(uv, \tau)) - \pi_t(u, \tau) \geq 0$ for all edges $uv \in E$ and times $\tau > \dist_w(s,u,\tdep)$.
        A* can be analyzed as an equivalent run of Dijkstra's algorithm with a modified weight function derived from the input weights and the potentials.
        With feasibility, these modified weights are non-negative, which implies correctness and polynomial running time.
        When $\pi_t(t, \tau) = 0$, feasibility also implies the lower bound property.
        However, feasibility is not strictly necessary to guarantee correctness.
  \item Lower bound: $\pi_t(v, \tau) \leq \dist_w(v,t,\tau)$ for every vertex $v \in V$ and time $\tau = \dist_w(s,v,\tdep)$.
        This ensures that the search has found the correct distance once the target vertex is settled.
        This is also sufficient for correctness.
        However, without feasibility, A* may settle vertices multiple times.
        In theory, this can lead to an exponential running time.
\end{itemize}

In Appendix~\ref{sec:appendix:correctness} we discuss these properties in detail and prove the correctness.
Note that these properties only need to hold for specific times $\tau$, not all possible times of the day.
Our practical potentials heavily rely on this and only compute data for the specific times necessary to answer a query correctly.

In the following, we present two practical realizations of time-dependent A* potentials.
Both are extensions of Lazy RPHAST.
Lazy RPHAST/CH-Potentials is already a very efficient potential and obtains exact distances for scalar lower bound weights, i.e.\ the tightest possible estimates with a time-independent potential definition.
To outperform CH-Potentials, on the one hand, we have to obtain significantly tighter estimates.
On the other hand, we also must avoid the potential evaluation becoming too expensive.
Therefore, we avoid costly operations on functions and work with scalar values as much as possible.
As a result, even though our potentials are time-dependent, computed estimates during a single query usually will \emph{not} change depending on the visit time of a vertex.

\subsection{Multi-Metric Potentials}

Let $(s,t,\tdep)$ be a query and $\tmax$ an upper bound on the optimal arrival time at the target.
Consider any $\tau' \leq \tdep$, $\tmax \leq \tau''$ and the weight function $l[\tau', \tau''](e) := \min_{\tau' \leq \tau \leq \tau''}\pred(e, \tau)$. 
Clearly, $\dist_{l[\tau', \tau'']}(v,t)$ provides lower bound estimates of distances to the target vertex during the time relevant for this query.
If $\tau'$ and $\tau''$ are close to $\tdep$ and $\tmax$ and, if the difference between $\tau'$ and $\tau''$ is not too big, the estimates will be significantly tighter than global lower bound distances.
The \emph{Multi-Metric Potentials} (MMP) approach is based on this observation.
Instead of using a single potential based on a global lower bound valid for the entire time, we process multiple lower bound weight functions for different time intervals.
At query time, we then select an appropriate weight function.
The upper bound $\tmax$ is computed with a time-independent CCH query on a scalar upper bound function $\comb^+_{\max}$ computed during the update phase.
Efficiently computing distances with respect to the selected weight function is done with Lazy RPHAST.
Therefore, no time-dependent computations need to be performed to evaluate this potential function.
MMP only depend on the departure time of the query but not of the potential evaluation time.
Still, MMP will be significantly tighter than any time-independent potential can be.

\subparagraph{Phase Details.}
The first step of the preprocessing for this potential is to perform the regular CCH preprocessing, i.e.\ compute an importance ordering and construct the unweighted augmented graph.
Now let $I$ be a set of time intervals.
In our implementation,
we cover the time between 6:00 and 22:00 with intervals of a length of one, two, four, and eight hours, starting every 30 minutes,
and one interval covering the entire day.
We do not maintain any additional intervals between 22:00 and 6:00 as most edge weights correspond to their respective free-flow travel time during this period.
Thus, the lower bound weights would be equal to the full-day lower bounds.
During preprocessing, for each interval $[\tau_i', \tau_i''] \in I$, we extract lower bound functions $l[\tau_i', \tau_i'']$ and run the CCH customization algorithm to obtain $l[\tau_i', \tau_i'']^+$.
This can be parallelized trivially.
Also, the customization can be parallelized internally.
For further engineering details, we refer to~\cite{dsw-cch-15,bsw-rttau-19,ghuw-fbndocch-19}.

During the update phase, we compute an additional lower bound weight function starting at $\tnow$ with duration $\delta$ derived from the combined weights $\comb$ and run the basic customization for it.
We use $\delta =$ 59 minutes to reasonably cover the live traffic but keep the live interval shorter than any other interval.
Further, we extract an upper bound weight function $\comb_{\max}$ which is valid for the entire day for both the predicted and the live traffic, and perform the CCH basic customization to obtain $\comb^+_{\max}$.

The query starts with a classical CCH query on the customized upper bound $\comb^+_{\max}$ to obtain a pessimistic estimate of $\tmax$.
We then select the smallest interval $[\tau_i', \tau_i'']$ such that $[\tdep,\tmax] \subseteq [\tau_i', \tau_i'']$.
Running Lazy RPHAST on $G^+_l$ with the customized weight function $l[\tau_i', \tau_i'']^+$ yields the desired potential function.
See Appendix~\ref{sec:appendix:optimizations} for additional optimizations.

\subparagraph{Correctness.}
For any given single query, the estimates obtained by MMP are actually time-independent.
They return the exact shortest distances with respect to a lower bound weight function valid for the query.
Constant potentials trivially adhere to the strong FIFO property.
Also, shortest distances for a lower bound weight function are feasible potentials~\cite{strasser_et_al:LIPIcs.SEA.2021.6}.

\subsection{Interval-Minimum Potentials}

\emph{Interval-Minimum Potentials} (IMP) is a time-dependent adaptation of the Lazy RPHAST algorithm.
While Lazy RPHAST has a single scalar weight for each edge, the Interval-Minimum Potential uses a time-dependent function.
This allows for tighter estimates but introduces new challenges.
First, we need an augmented graph with sufficiently accurate time-dependent lower bounds.
We utilize the existing CATCHUp customization~\cite{swz-sfert-21} because it is based on CCH.
Second, storing the shortcut travel time functions $w^+$ may consume a lot of memory.
Further, the representation as a list of breakpoints makes the evaluation more expensive than looking up a scalar weight.
Therefore, we resort to a different representation and store functions as piecewise constant values in buckets of equal duration.
Third, evaluating these functions requires a time argument.
While $\pi_t(v, \tau)$ includes the time argument $\tau$ for the time at $v$, Lazy RPHAST also needs a time for every recursive invocation.
Therefore, we apply Lazy RPHAST a second time on global upper and lower bound weight functions $\comb_{\max}^+$ and $\pred_{\min}^+$ to quickly obtain arrival intervals for arbitrary vertices.
We then use these intervals to evaluate the edge weights and obtain tight time-dependent lower bounds.

\subparagraph{Phase Details.}
The first preprocessing step is the CCH preprocessing.
For the second step, we need to obtain time-dependent travel times for the augmented graph $G^+$ based on the predicted traffic weights $p$.
For this, we utilize CATCHUp~\cite{swz-sfert-21}, a time-dependent adaptation of CCH.
The CATCHUp customization yields for each edge in $uv \in E^+$ approximated \emph{time-dependent} lower bound functions $\underline{b^+}(uv)$.
We transform the time-dependent piecewise linear lower bound functions $\underline{b^+}$ into piecewise \emph{constant} lower bound functions
$\pcfn(uv, \tau) := \min \left\{ \underline{b^+}(uv, \tau') \mid \bucketlen  \lfloor \frac{\tau}{\bucketlen} \rfloor \leq \tau' < \bucketlen (\lfloor \frac{\tau}{\bucketlen} \rfloor + 1) \right\}$
where $\bucketlen$ is the length of each constant segment.
This enables a compact representation.
Functions can be represented with a fixed number of values per edge.
We use 96 buckets of length $\bucketlen = 15$ minutes.
Additionally, we derive a scalar lower bound $b^+_{\min}$.
Note that $b^+_{\min}$ is typically tighter than bounds obtained by a time-independent customization on lower bounds of the input functions, i.e.\ $w^+_{\min}$.

In the update phase, we extract a combined traffic upper bound weight function $\comb_{\max}$ for the entire day and run the CCH customization to obtain $\comb^+_{\max}$.

The query consists of two instantiations of the Lazy RPHAST algorithm.
The first one uses the scalar bounds $b^+_{\min}$ and $\comb^+_{\max}$ and computes an interval of possible arrival times at arbitrary vertices when departing from $s$ at $\tdep$.
Since arrival intervals are distances from the source vertex, we have to apply Lazy RPHAST in reverse direction.
This means we first run Dijkstra's algorithm from $s$ on $\gchu$, and then, we apply the recursive distance-memoizing DFS on $\gchd$ for any vertex for which we want to obtain an arrival interval.
We denote this instance as AILR for \emph{Arrival Interval Lazy RPAHST}.
With these arrival intervals, we can now compute lower bounds to the target with the second Lazy RPHAST instantiation, which uses the time-dependent lower bounds $\pcfn$.
The first step is to run Dijkstra's algorithm from $t$ on $\gchd$.
To relax an edge $uv \in \rechd$, we first need to obtain an arrival interval $[\tau_{\min}, \tau_{\max}]$ at $v$ using AILR.
This allows us to determine for $vu$ at the relevant time a tight lower bound $d := \min_{\tau \in [\tau_{\min}, \tau_{\max}]} \pcfn(vu, \tau)$.
Then, we check if we can improve the lower bound from $v$ to $t$, i.e.\ $\mathtt{D}[v] \leftarrow \min(\mathtt{D}^{\downarrow}[v], \mathtt{D}[u] + d)$.
Having established preliminary backward distances for all vertices in the CH search space of $t$, we can now compute estimates with the recursive distance-memoizing DFS.
To obtain a distance estimate for vertex $u$, we first recursively compute distance estimates $\mathtt{D}[v]$ for all upward neighbors $v$ where $uv \in \echu$.
Then, we use AILR to obtain an arrival interval $[\tau_{\min}, \tau_{\max}]$ at $u$.
Finally, we relax the upward edges $uv$ set $\mathtt{D}[u] \leftarrow \min(\mathtt{D}^{\downarrow}[u], \mathtt{D}[v] + \min_{\tau \in [\tau_{\min}, \tau_{\max}]} \pcfn(uv, \tau))$.
This yields the final estimate for $u$.

Choosing a good memory layout for the bucket weights is crucial for the performance.
We store all edge weights of each bucket consecutively.
Typically, only a few buckets per edge are relevant because the arrival intervals are relatively small.
Also, all outgoing edges of each vertex are evaluated consecutively.
Thus, having their weights for the same bucket close to each other increases cache hits.
See Appendix~\ref{sec:appendix:optimizations} for additional optimizations.

\subparagraph{Correctness.}
Estimates obtained by IMP are lower bounds of the actual time-dependent shortest distances.
This directly follows from the correctness of the CATCHUp preprocessing and the Lazy RPHAST algorithm.
Also, they do satisfy the strong FIFO property because, for any given single query, the estimates are constant.
However, they are not feasible due to the piecewise constant approximation schema.
We could not observe any practical negative consequences of this, though.

\subsection{Compression}\label{sec:compression}

Both of our time-dependent potentials use many weight functions.
This can lead to problematic memory consumption.
However, since we only need lower bounds, we can merge weight functions.
Consider two MMP intervals with weight functions $l_1$ and $l_2$.
A combined function $l_{1 \cup 2}(uv) = \min(l_1(uv), l_2(uv))$ is valid for both intervals, albeit less tight.
We can merge IMP buckets analogously.
Thus, we can reduce memory consumption by trading tightness.
Both potentials can handle merged lower bound functions with a layer of indirection:
Buckets and intervals are mapped to a weight function ID.
The weight of an edge in a merged weight function is the minimum weight of this edge in all included functions.

We now discuss an efficient and well-parallelizable algorithm to iteratively merge weight functions until only $k$ functions remain.
In each step, we merge the pair of weight functions with the minimal sum of squared differences of all edge weights.
Since comparing all pairs of weight functions is expensive, we track the minimum difference sum $\Delta_{\min}$ we have found so far and stop any comparison where the sum exceeds $\Delta_{\min}$.
However, even when stopping a comparison, we store the preliminary sum and the edge ID up to which we have summed up the differences.
Then, we do not need to start from scratch should we continue to compare this particular pair of weight functions.
Finally, we maintain all pairs of weights along with the (possibly preliminary) difference sums in a priority queue ordered by the difference sums.
When merging two weight functions, all other associated queue entries are removed from the queue and new entries for comparisons between the new weight function and all other functions are inserted.
To determine the next weight function pair to merge, unfinished weight function pairs are popped from the queue and processed in parallel.
The minimum difference is tracked in an atomic variable.

\section{Evaluation}
\subparagraph{Environment.} Our benchmark machine runs openSUSE Leap 15.3 (kernel 5.3.18), and has 192\,GiB of DDR4-2666 RAM and two Intel Xeon Gold 6144 CPUs, each of which has 8 cores clocked at 3.5\,GHz and 8~$\times$~64\,KiB of L1, 8~$\times$~1\,MiB of L2, and 24.75\,MiB of shared L3 cache.
Hyperthreading was disabled and parallel experiments use 16 threads.
We implemented our algorithms in Rust\footnote{
Our code and experiment scripts are available at \url{https://github.com/kit-algo/tdpot}
} and compiled them with \texttt{rustc 1.61.0-nightly (c84f39e6c 2022-03-20)} in the release profile with the \texttt{target-cpu=native} option.

\subparagraph{Datasets.}
We evaluate our algorithms on two networks for which we have proprietary traffic data available.
Sadly, we cannot provide access to these datasets due to non-disclosure agreements.
We are not aware of any publicly available real-world traffic feeds or predictions.
However, as these datasets are the same ones used in~\cite{swz-sfert-21,strasser_et_al:LIPIcs.SEA.2021.6}, at least some comparability is given.
Our first network, PTV Europe, has been provided by PTV\footnote{\url{https://ptvgroup.com}} in 2020 and is based on TomTom\footnote{\url{https://www.tomtom.com}} routing data covering Western Europe.
It has 28.5\,M vertices and 61\,M edges.
76\% of the edges have a non-constant travel time.
The data includes a traffic incident snapshot from 2020/10/28 07:47 with live speeds and estimated incident durations for 215\,k vertex pairs.
Our second network, OSM Germany, is derived from an early 2020 snapshot of OpenStreetMap and was converted into a routing graph using RoutingKit\footnote{\url{https://github.com/RoutingKit/RoutingKit}}.
It has 16.2\,M vertices and 35.2\,M edges.
For this instance, we have proprietary traffic data provided by Mapbox\footnote{\url{https://mapbox.com}}.
This includes traffic predictions for 38\% of the edges in the form of predicted speeds for all five-minute periods over the course of a week.
We only use the predictions for one day.
Also, we exclude speed values which are faster than the free-flow speed computed by RoutingKit.
The data also includes two live traffic snapshots in the form of OSM node ID pairs and live speeds for the edge between the vertices.
One is from Friday 2019/08/02 15:41 and contains 320\,k vertex pairs and the other from Tuesday 2019/07/16 10:21 and contains 185\,k vertex pairs.
The datasets do not contain any estimate for how long the observed live speeds will be valid.
We set $\tend$ to one hour after the snapshot time.
Note that even though it is smaller and has fewer time-dependent edges, OSM Germany is actually the harder instance.
This is because it has more breakpoints per time-dependent edge (124.8 compared to 22.5 on PTV Europe) and the predicted travel times fluctuate more strongly.

\subparagraph{Methodology.}
We evaluate our algorithms by sequentially solving batches of 100\,k shortest path queries with three different query sets:
First, there are \emph{random} queries where source and target are drawn from all vertices uniformly at random.
These are mostly long-range queries.
Second are \emph{1h} queries where we draw a source vertex uniformly at random, run Dijkstra's algorithm from it and pick the first node with a distance greater than one hour as the target.
Third, we generate queries following the Dijkstra rank methodology~\cite{ss-hhhes-05} to investigate the performance with respect to query distance.
For these \emph{rank} queries, we pick a source uniformly at random and run Dijkstra's algorithm from it.
We use every $2^{i}$-th settled vertex as the target for a query of Dijkstra rank $2^i$.
For queries with only predicted traffic, we pick $\tdep$ uniformly at random.
When using live traffic, we set $\tdep = \tnow$.
To evaluate the performance of the preprocessing and update phases, we run them 10 and 100 times, respectively.
Preprocessing and update phases utilize all cores using 16 threads.

We compare our time-dependent potentials MMP and IMP against time-independent CH-Potentials algorithm realized on CCH.
Therefore, we denote this approach as CCH-Potentials.
All three potentials use the same CCH vertex order and augmented graph.
CCH-Potentials provide heuristic estimates based on a lower bound without any real-time or predicted traffic.
Thus, no update phase is necessary to integrate real-time traffic updates.
It is the only other speedup technique we are aware of that supports exact queries for our problem model.
Dijkstra's algorithm without any acceleration is our baseline.

\begin{table}
\centering
\caption{
Query and preprocessing performance results of different potential functions on different graphs and live traffic scenarios.
We report average running times, number of queue pops, relative increases of the result distance over the initial distance estimate and speedups over Dijkstra's algorithm for 100\,k random queries.
Additionally, we report preprocessing and update times and the memory consumption of precomputed auxiliary data.
}\label{tab:pot_perf}
\begin{tabular}{cccrrrrrrr}
\toprule
 &        &    Live & Running   & Queue          &  Length     & Speedup & Prepro. & Update & Space \\
 & Graph  & traffic & time [ms] & [$\cdot 10^3$] &  incr. [\%] &         &     [s] &    [s] &  [GB] \\
\midrule
\multirow{5}{*}{\rotatebox[origin=c]{90}{CCH Pot.}} & \multirow{3}{*}{Ger} &    -- &     137.5 &           92.3 &        12.2 &    24.8 &                            &    -- &  0.8 \\ [-1.5pt]
                                                    &                      & 10:21 &     236.5 &          158.3 &        18.9 &    14.7 &                      165.2 &    -- &  0.8 \\ [-1.5pt]
                                                    &                      & 15:41 &     128.0 &           89.6 &        19.1 &    27.0 &                            &    -- &  0.8 \\ [2pt]
                                                    & \multirow{2}{*}{Eur} &    -- &     102.6 &           65.2 &         4.2 &    58.0 &     \multirow{2}{*}{249.7} &    -- &  1.0 \\ [-1.5pt]
                                                    &                      & 07:47 &     152.2 &          102.2 &         8.4 &    39.3 &                            &    -- &  1.0 \\ \addlinespace
\multirow{5}{*}{\rotatebox[origin=c]{90}{MMP}}      & \multirow{3}{*}{Ger} &    -- &     117.7 &           74.6 &         9.9 &    29.0 &                            &    -- & 33.7 \\ [-1.5pt]
                                                    &                      & 10:21 &     170.0 &          110.0 &        13.0 &    20.4 &                      382.6 &  15.2 & 34.0 \\ [-1.5pt]
                                                    &                      & 15:41 &     119.0 &           79.5 &        15.8 &    29.0 &                            &  15.3 & 34.0 \\ [2pt]
                                                    & \multirow{2}{*}{Eur} &    -- &      95.3 &           58.6 &         3.5 &    62.5 &     \multirow{2}{*}{581.5} &    -- & 56.2 \\ [-1.5pt]
                                                    &                      & 07:47 &     131.2 &           84.5 &         5.8 &    45.6 &                            &  22.7 & 57.2 \\ \addlinespace
\multirow{5}{*}{\rotatebox[origin=c]{90}{IMP}}      & \multirow{3}{*}{Ger} &    -- &      22.2 &            5.1 &         1.8 &   154.1 &                            &    -- & 30.7 \\ [-1.5pt]
                                                    &                      & 10:21 &      29.1 &            7.6 &         2.6 &   119.2 &                  13\,687.0 &  13.5 & 31.2 \\ [-1.5pt]
                                                    &                      & 15:41 &      37.7 &           11.3 &         4.2 &    91.5 &                            &  13.6 & 31.2 \\ [2pt]
                                                    & \multirow{2}{*}{Eur} &    -- &      11.5 &            1.8 &         0.4 &   518.0 &  \multirow{2}{*}{1\,799.9} &    -- & 52.1 \\ [-1.5pt]
                                                    &                      & 07:47 &      25.4 &            7.4 &         1.7 &   235.5 &                            &  20.1 & 53.1 \\
\bottomrule
\end{tabular}

\end{table}

\subparagraph{Experiments.}
In Table~\ref{tab:pot_perf}, we report key performance results for our time-dependent potentials on random queries.
We observe that IMP is the fastest approach by a significant margin, up to an order of magnitude faster than time-independent CCH-Potentials and roughly two orders of magnitude faster than Dijkstra's algorithm.
The search space reduction is even greater, but this does not fully translate to running times due to the higher evaluation overhead of IMP.
With only predicted traffic, IMP is only two to three times slower than CATCHUp~\cite{swz-sfert-21}.
This shows that using A* to gain algorithmic flexibility comes at a price, but the overhead compared to purely hierarchical techniques is manageable.
In contrast, MMP is only slightly faster than CCH-Potentials.
This is expected since random queries are mostly long-range for which MMP is not particularly well suited.

\begin{figure}[!t]
\centering
\includegraphics[width=\linewidth]{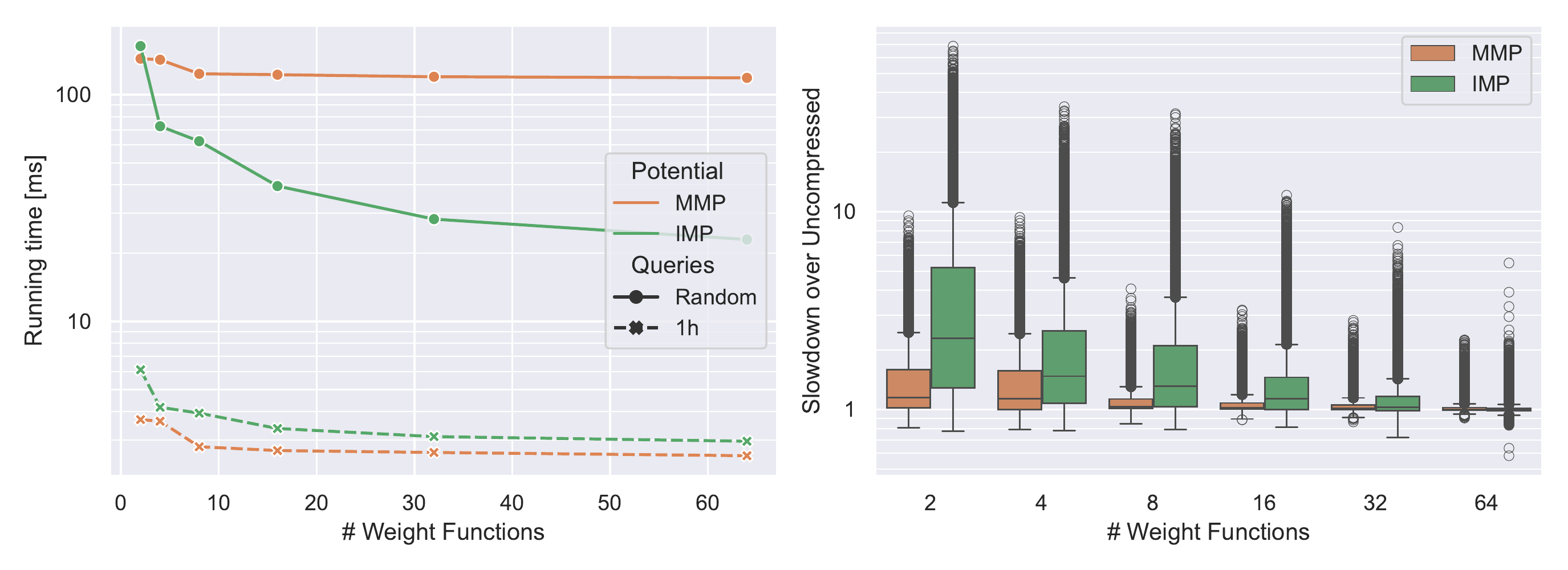}
\caption{
Left: Mean running times of 100\,k queries on OSM Germany with only predicted traffic by number of remaining weight functions.
Right: Boxplot of the per-query relative slowdown over the running time of the respective query with all weight functions.
}\label{fig:compression}
\end{figure}

Preprocessing times are within a couple of minutes for CCH-Potentials and MMP.
IMP preprocessing is significantly more expensive because of the time-dependent CATCHUp preprocessing.
This is especially pronounced on OSM Germany where the time-dependent travel time functions fluctuate strongly.
Still, running preprocessing algorithms on a daily basis is quite possible.
This also underlines that frequently running a CATCHUp customization to include live traffic is not feasible.
For both our approaches, real-time traffic updates are possible within a fraction of a minute.
MMP is slightly slower because it uses a few more weight functions.
Both our approaches are quite expensive in terms of memory consumption, but this can be mitigated through the use of compression (see Figure~\ref{fig:compression}).

Introducing live traffic decreases the quality of the estimates and thus increases search space sizes and running times.
For IMP, this increases running times by roughly a factor of two.
Even with heavy rush hour traffic, IMP is still more than 90 times faster than Dijkstra's algorithm.
Surprisingly, for CCH-Potentials and MMP, this scenario seems easier to handle than light midday traffic.
This actually is an effect of the \emph{predicted} traffic.
It also has a strong influence on the performance of CCH-Potentials and MMP depending on the departure time.

\begin{figure}[tb]
\centering
\includegraphics[width=\linewidth]{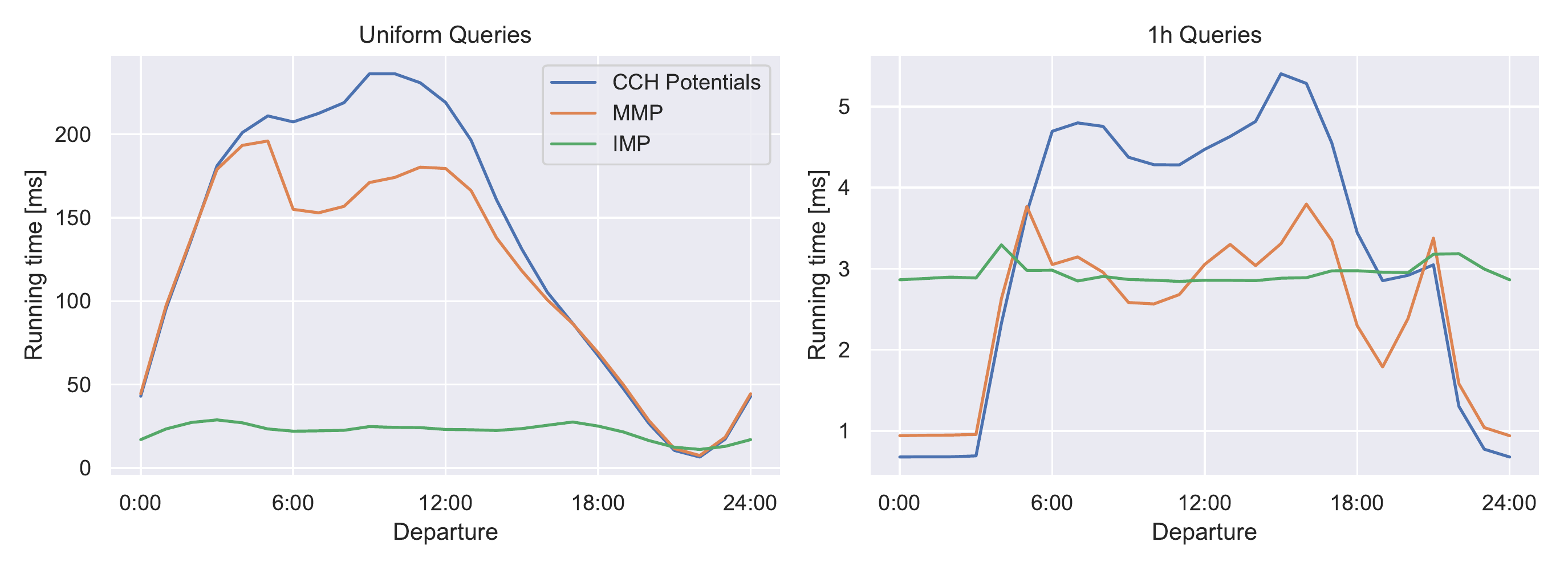}
\caption{
Average running time of 100\,k uniform and 1h queries on OSM Germany with only predicted traffic.
Each query has a departure time drawn uniformly at random.
The resulting running times are grouped by the departure time hour.
}\label{fig:by_dep}
\end{figure}

We investigate this behavior with Figure~\ref{fig:by_dep} which depicts query performance by departure time over the course of the day.
Clearly, the departure time has a significant influence both for short-range and long-range queries.
For long-range queries, the peaks are shifted and smeared because of the travel time (4--5 hours on average on OSM Germany) covered by the query.
This is the reason why the heavy afternoon traffic appears to be easier than the light midday traffic for MMP and CCH-Potentials.
For IMP, the influence of the departure time is much smaller, which makes it consistently the fastest approach on long-range queries.
For short-range queries, the overhead of IMP make it the slowest during the night.
Moreover, MMP is roughly as fast as IMP for 1h queries during the daytime.
Therefore, MMP may actually be a simple and effective approach for practical applications where short-range queries are more prominent.

\begin{figure}[!b]
\centering
\includegraphics[width=\linewidth]{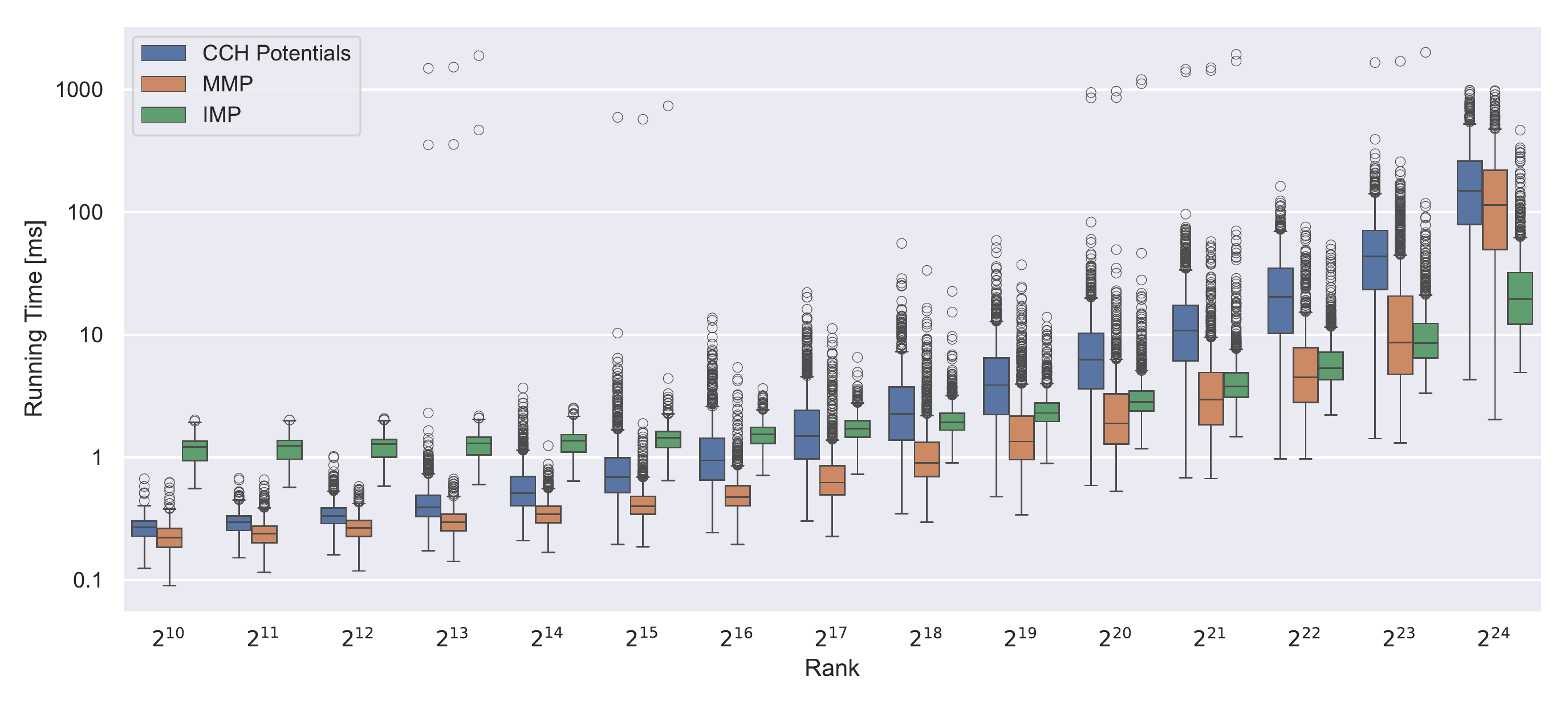}
\caption{
Box plot of running times for 1\,000 queries per Dijkstra-rank on PTV Europe with live traffic and fixed departure at 07:47.
The boxes cover the range between the first and third quartile.
The band in the box indicates the median; the whiskers cover 1.5 times the interquartile range.
All other running times are indicated as outliers.
}\label{fig:ranks}
\end{figure}

Figure~\ref{fig:ranks} depicts the performance by query distance.
For short-range queries, IMP is slower than the other approaches because the potential is expensive to evaluate, but it scales much better to long-range queries because of its estimates are tighter.
Also, the variance in running times is significantly smaller.
Even for rank $2^{24}$, most queries can be answered within a few tens of milliseconds.
Nevertheless, MMP is actually faster on most ranks.
Only at rank $2^{24}$, MMP running times become as slow as the CCH-Potentials baseline.
A jump in MMP running times can be observed from rank $2^{23}$ to $2^{24}$.
This is because the mean query distance jumps from five to six hours on rank $2^{23}$ to over eight hours on rank $2^{24}$, which is longer than the longest covered interval.
Thus, on rank $2^{24}$, MMP fall back to classical CCH-Potentials on many queries.
We also observe a few strong outliers.
This happens because of blocked streets in the live traffic data.
When the target vertex of a query is only reachable through a blocked road segment, A* will traverse large parts of the networks until the blocked road opens up.
This affects all three potentials in the same way and demonstrates an inherent weakness of A*-based approaches: the performance always depends on the quality of the estimates.
However, on realistic instances, the time-dependent preprocessing algorithms of purely hierarchical approaches are too expensive for frequent rerunning.
This makes our approach the first to enable interactive query times across all distances in a setting with combined live and predicted traffic.

Finally, Figure~\ref{fig:compression} showcases the effects of reducing the number of weight functions.
MMP appears to be very robust against compression.
We can reduce the number of weight functions to 16 (a memory usage reduction of about a factor of 6) before the slowdowns become noticeable in the mean running time.
However, MMP only achieves relatively small speedups compared to CCH-Potentials, i.e.\ rarely more than a factor of three.
Therefore, its robustness is not particularly surprising.
IMP, which achieves stronger speedups, is less robust against compression.
Nevertheless, we can reduce the memory consumption by a factor of about three to 32 functions and still achieve very decent query times.
With 32 functions, the absolute memory consumption decreases to less than 20\,GB, which is at least manageable.
Surprisingly, even with only four weight functions, IMP is still faster than MMP on long-range queries.
This clearly shows the superiority of IMP for long-range queries.
The compression algorithm itself takes less than a minute, depending on the final number of weight functions.
Thus, its running time is dominated by the regular preprocessing.
See Appendix~\ref{sec:appendix:parallelization} for further details on the effectiveness of the parallelization.

\section{Conclusion}

In this paper, we proposed time-dependent A* potentials for efficient and exact routing in time-dependent road networks with both predicted and live traffic.
We presented two realizations of time-dependent potentials with different trade-offs.
Both allow fast live traffic updates within a fraction of a minute.
IMP achieves query times two orders of magnitude faster than Dijkstra's algorithm and up to an order of magnitude faster than state-of-the-art time-independent potentials.
To the best of our knowledge, this makes our approach the first to achieve interactive query performance while allowing fast updates in this setting.
For future work, we would like to apply our time-dependent potentials to other extended scenarios in time-dependent routing, for example to incorporate turn costs.

\appendix

\section{Further Discussion of Traffic Model}\label{sec:appendix:model}

In~\cite{dn-crdtd-12}, ALT-based algorithms supporting dynamic updates to $\pred$ are discussed.
This might seem more flexible than our model.
However, we believe that our model is, in fact, more practical.
The problem with live traffic updates to $\pred$ is that these predictions are assumed to be periodic.
But live traffic incidents are inherently tied to the current moment and are not expected to repeat in 24 hours.

To the best of our knowledge, the first model for combined traffic separating predictions and real-time information was proposed in~\cite{strasser:OASIcs:2017:7897}.
This approach was also employed in~\cite{strasser_et_al:LIPIcs.SEA.2021.6} and is quite similar to ours.
Our model only slightly differs in two ways.
First, in our model, the combined function equals the predicted function from moment $\tend(e)$ on, while in~\cite{strasser:OASIcs:2017:7897}, the combined function has the live weight until $\tend(e)$ and only then starts to approach the predicted function again.
This allows us to handle blocked edges more realistically.
In the case of $\live(e) = \infty$, i.e.\ the edge is blocked, our combined travel time function for the edge corresponds to waiting until the edge opens up at $\tend(e)$.
With the model from~\cite{strasser:OASIcs:2017:7897}, the combined function would decrease from $\infty$ towards the predicted function but would never reach it and thus stay completely blocked.
Second, in our model, we use $\max(\pred(e, \tau), \live(e))$, while the authors of~\cite{strasser:OASIcs:2017:7897} use $\max(\pred_{\min}(e), \live(e))$, i.e.\ our model does not allow the live travel time to be less than the predicted travel time.
While this is a significant restriction from a theoretical perspective, there are good reasons why this is a valid modeling assumption for practical purposes.
First, the assumption is realistic.
Live traffic should account for unexpected traffic events.
Therefore, live travel times will rarely be better than predicted travel times.
If the live traffic is frequently better than the predicted traffic, the predictions should be adjusted.
Second, a less flexible model which can be solved efficiently and exactly is often more practical than a more flexible one that can only be solved heuristically.
Further, the reason to solve routing problems to exactness is not that the computed travel times will perfectly match reality.
Traffic data is always only an approximation of reality.
Rather, obtaining exact results helps to present users with consistent routes where travel times do not suddenly become better when the user takes an unexpected alternative.
Further, it allows attributing ``weird'' routes to either bugs in the data or bugs in the implementation which is very useful in practice.

\section{Correctness Properties of Time-Dependent A* Potentials}\label{sec:appendix:correctness}

In this section, we analyze correctness properties for time-dependent A* potentials.
Here, it is often more practical to use arrival time functions instead of travel time functions.
Given a travel time function $w$, we denote the respective arrival time function as $\hat{w}(uv, \tau) := w(uv, \tau) + \tau$.
With arrival time functions, we can represent path lengths simply as the composition of the arrival time functions of the individual edges.
We use $(f \circ g)(x) = f(g(x))$ to denote the function composition.

Similarly to the time-independent case, we can define a modified weight function $w'$ such that running A* on the graph $G$ with weights $w$ with a time-dependent potential $\pi_t$ is equivalent to running Dijkstra's algorithm on $G$ with modified weights $w'$.
Consider a vertex $u$ with an arrival time of $\tau$.
In A*, its queue key is $\tau' = \pi_t(u, \tau) + \tau = \hat{\pi}_t(u, \tau)$.
For the equivalent run of Dijkstra's algorithm, we want the distances to equal these queue keys.
Therefore, the reduced weight function of edge $uv$ must compose the original function with the potential at the head vertex $v$, i.e.\ $\hat{\pi}_t(v, \hat{w}(uv, \tau)) = (\hat{\pi}_t(v) \circ \hat{w}(uv))(\tau)$.
However, the input of $w$ is a time $\tau$, not the A* queue key $\tau' = \hat{\pi}_t(u, \tau)$.
To first go back from $\tau'$ to $\tau$, we also need to compose this with the inverted potential function $\hat{\pi}_t(u)^{-1}$.
Thus, the modified weight $\hat{w}'$ are defined as $\hat{w}'(uv, \tau') = \hat{\pi}_t(v, \hat{w}(uv, \hat{\pi}_t(u)^{-1}(\tau'))) = (\hat{\pi}_t(v) \circ \hat{w}(uv) \circ \hat{\pi}_t(u)^{-1})(\tau')$.

For $\hat{\pi}_t(u)^{-1}$ to be well-defined, we need $\tau_1 \neq \tau_2 \implies \hat{\pi}_t(\tau_1) \neq \hat{\pi}_t(\tau_2)$.
Therefore, for potentials, we require the \emph{strong First-In First-Out} property:
\[
\pi_t(v, \tau) < \pi_t(v, \tau + \epsilon) + \epsilon, v \in V, \epsilon > 0
\]
Note that potentials that only adhere to the regular FIFO property might also work in practice.
However, in this case, the modified weights for a theoretically equivalent run of Dijkstra's algorithm are not well-defined anymore.
This breaks the following correctness argument.
Nevertheless, a more sophisticated analysis could probably work around this problem.

Shortest paths for the modified weights $w'$ are the same as with the original weights $w$.
Consider a path $P = (s,v_1,\dots,v_k,t)$.
By definition, the arrival time function of a path is
\[
\hat{w}'(P) = \hat{\pi}_t(t) \circ \hat{w}(v_k t) \circ \hat{\pi}_t(v_k)^{-1}
\circ \hat{\pi}_t(v_k) \circ \dots \circ \hat{\pi}_t(v_1)^{-1} \circ
\hat{\pi}_t(v_1) \circ \hat{w}(s v_1) \circ \hat{\pi}_t(s)^{-1}
\]
Note that all the inner potential evaluations cancel out.
Besides the initial and final potential evaluations, only the composition of the original weights remains.
Now consider the modified weights of two different $s$-$t$ paths.
With a fixed departure, the initial inverted potential at $s$ is the same for both paths.
Thus, the original weights determine which path is shorter.
The final potential evaluation at $t$ cannot change the relative order due to the FIFO property.
Therefore, shortest paths for $w'$ are the same as for $w$.

We can infer that if these modified weights do not have any negative travel times, then Dijkstra and thus A* will obtain correct results. 
This leads to the \emph{feasibility} property for time-dependent potentials:
\[
\hat{w}'(uv, \tau') = \hat{\pi}_t(v, \hat{w}(uv, \hat{\pi}_t(u)^{-1}(\tau'))) \geq \tau'
\]
To simplify correctness proofs for potentials, we reformulate this to a simpler equivalent property:
\begin{align*}
\hat{\pi}_t(v, \hat{w}(uv, \tau)) & \geq \hat{\pi}_t(u, \tau) \\
\pi_t(v, w(uv, \tau) + \tau) + w(uv, \tau) + \tau & \geq \pi_t(u, \tau) + \tau \\
\pi_t(v, w(uv, \tau) + \tau) + w(uv, \tau) - \pi_t(u, \tau) & \geq 0
\end{align*}
This formulation also shows that the time-dependent feasibility is a generalization of the classical feasibility property $w(uv) + \pi_t(v) - \pi_t(u) \geq 0$.

The third property is the \emph{lower bound} property:
\[
\pi_t(v, \tau) \leq \dist_w(v,t,\tau)
\]
With a potential fulfilling this property, A* is guaranteed to have found the shortest path once the target vertex is settled, even if the potential is not feasible.
This is because every vertex on the shortest path must have a queue key less or equal to the target when discovered with optimal distance.
Suppose for contradiction the target vertex would be settled with distance $d$ which is greater than the shortest distance.
But because of the lower bound property, every vertex on the shortest path will have a queue key smaller than $d$ as soon as it was discovered with its shortest distance.
By induction, this must happen for all vertices on the shortest path.
Thus, also the shortest path to $t$ must have been found.
This contradicts our assumption.

Also, even with negative travel times due to infeasible potentials, negative cycles are not possible.
Consider a cycle $C$ starting and ending at vertex $v$.
The length of the cycle with the modified weights is $w'(C, \tau') = \hat{\pi}_t(v, \hat{w}(C, \hat{\pi}_t(u)^{-1}(\tau')))$ because all inner potential functions cancel out.
As $\hat{\pi}_t(v)$ has to adhere to the FIFO property, the length of $C$ cannot be negative.
We conclude that the distance at $t$ is final as soon as $t$ is settled.
Thus, A* with lower bound potentials will obtain correct results.
However, the running time may be exponential in the graph size.

Crucially, to guarantee correctness, potentials need not adhere to these properties for all points in time $\tau$.
Assume we are running A* to answer a query from $s$ to $t$ with departure $\tdep$.
Clearly, A* will never invoke the potential $\pi_t(v, \tau)$ of a vertex $v$ with $\tau < \dist(s,v,\tdep)$.
Therefore, it is sufficient to guarantee the strong FIFO property at vertex $v$ and the feasibility property for edge $vw$ for $\tau \geq \dist(s,v,\tdep)$.
For the lower bound property it even suffices to guarantee it only at exactly $\tau = \dist(s,v,\tdep)$.
When a vertex is discovered with a suboptimal distance, the queue key may be arbitrarily larger but never smaller because of the FIFO property.
Therefore, the inductive argument still applies.
All vertices on the shortest path will be traversed before the target is settled.

\section{Optimizations}\label{sec:appendix:optimizations}

\subsection{Perfect Customization}

Recall that shortcut edges $uv \in E^+$ allow skipping over paths of lower-ranked vertices.
We now denote the length of the shortest such path by $\dist^{\prec}_w(u,v) = w^+(uv)$.
In the main part of the paper, we stated that the CCH customization yields the appropriate weight function $w^+$ for the full augmented graph $G^+$ with $w^+(uv) = \dist^{\prec}(u,v)$.
This is only the \emph{basic customization}.
Because the augmented graph $G^+$ is valid for all possible weight functions, it contains many unnecessary edges for any concrete weight function.
This can be addressed with two additional steps in the customization phase.
In the \emph{perfect customization}, edges $uv \in E^+$ are processed again to obtain a weight function $w^*(uv) = \dist(u,v)$.
The authors of~\cite{dsw-cch-15} proved that edges $uv$ where $w^+(uv) > w^*(uv)$ are not necessary to answer queries correctly.
Thus, in a final step, these edges are removed, and a \emph{reduced augmented graph} $G^*=(V, E^*)$ is constructed.
All three customization steps can be parallelized efficiently~\cite{bsw-rttau-19}.

CATCHUp does not support a time-dependent \emph{perfect} customization.
However, a decent number of unnecessary edges can be identified by running the time-independent perfect customization on $b^+_{\max}$ to obtain $b^*_{\max}$.
Each edge $uv$ where $\dist_w(u,v,\tau) \leq b^*_{\max}(uv) < b^+_{\min}(uv) \leq \dist^{\prec}_w(u,v,\tau)$ can be removed.
We utilize this insight for our time-dependent potentials.

During the MMP update phase, we also run the perfect customization for the upper bound and obtain $\comb^*_{\max}$.
This allows us to remove unnecessary edges and accelerate the query phase.
We remove edges $uv$ where $\dist^{\prec}_c(u,v) \geq l_{[0:00, 24:00]}^+(uv) > \comb^*_{\max}(uv) \geq \dist_c(u,v)$ for all weight functions.
We could remove even more edges if we were to check the condition for every lower bound weight function individually.
However, in this case, we could not reuse the same reduced augmented graph topology for all weight functions.
As this would roughly double memory consumption, we only remove edges that can be removed for all weight functions.
We parallelize the reduced graph constructions as described in~\cite{bsw-rttau-19}.

For IMP, we can employ the same technique.
We identify unnecessary edges in the update phase and construct a smaller reduced augmented graph.
This accelerates queries.
Again we use the perfect customization to obtain $\comb^*_{\max}$ during the update phase and remove edges $uv$ where $\comb^*_{\max}(uv) < b^+_{\min}(uv)$.

\subsection{Metric Switching}

The tightness of MMP can be improved further by switching to weight functions for smaller intervals as the query progresses.
When initializing the potential, we select, beside $[\tau_i', \tau_i''] \supseteq [\tdep,\tmax]$, additional intervals $[\tau_j', \tau_j''] \subset [\tau_i', \tau_i'']$ with $\tau_j'' \geq \tmax$.
When evaluating the potential of a vertex at instant $\tau$, we use the weight function with the shortest associated interval that still includes $\tau$.
However, running Lazy RPHAST separately for each of these weight functions would be too expensive.
Therefore, we modify the DFS part of Lazy RPHAST to track, besides the distances $\mathtt{D}[u]$, also the weight function $\mathtt{W}[u]$ which was used.
Then, an already memoized distance $\mathtt{D}[v]$ for vertex $v$ can be reused if the interval associated with $\mathtt{W}[v]$ is a superset of the best interval for $\tau$.
We pass the $\tau$ parameter unmodified to recursive invocations.
Thus, we maintain the invariant that if $\mathtt{D}[v]$ is final and valid for $\tau$, also all vertices in the CH search space of $v$ will have a valid lower bound for an interval that includes $\tau$.
Therefore, we will only underestimate lower bounds.
The lower bound property is preserved.
The strong FIFO property is preserved, too.
Potential functions are now piecewise constant with only increasing jumps.
However, this breaks the feasibility property.
We could not observe any practical negative consequences of this, though.

For IMP, we employ a similar approach.
We use the potential evaluation time argument $\tau$ to tighten the arrival intervals and reduce the number of buckets to look up.
When evaluating the potential of a vertex $v$ at instant $\tau$, we still obtain the arrival interval $[\tau_{\min}, \tau_{\max}]$ using the AILR.
Typically, $\tau$ will be greater than $\tau_{\min}$.
Therefore, we can avoid looking up unnecessary bucket entries by only evaluating $[\max(\tau_{\min}, \tau), \tau_{\max}]$.
We track the respective bucket of $\tau$ in $\mathtt{B}[v]$.
When the potential is evaluated again with a possibly smaller $\tau$, we check if the respective bucket is smaller than $\mathtt{B}[v]$ to determine if the memoized distance can be reused.
If not, we reevaluate outgoing edges with the additional buckets and update $\mathtt{B}[v]$ accordingly.
We pass the $\tau$ parameter unmodified to recursive invocations.

\section{Parallelization of Compression}\label{sec:appendix:parallelization}

\begin{figure}
\centering
\includegraphics[width=\linewidth]{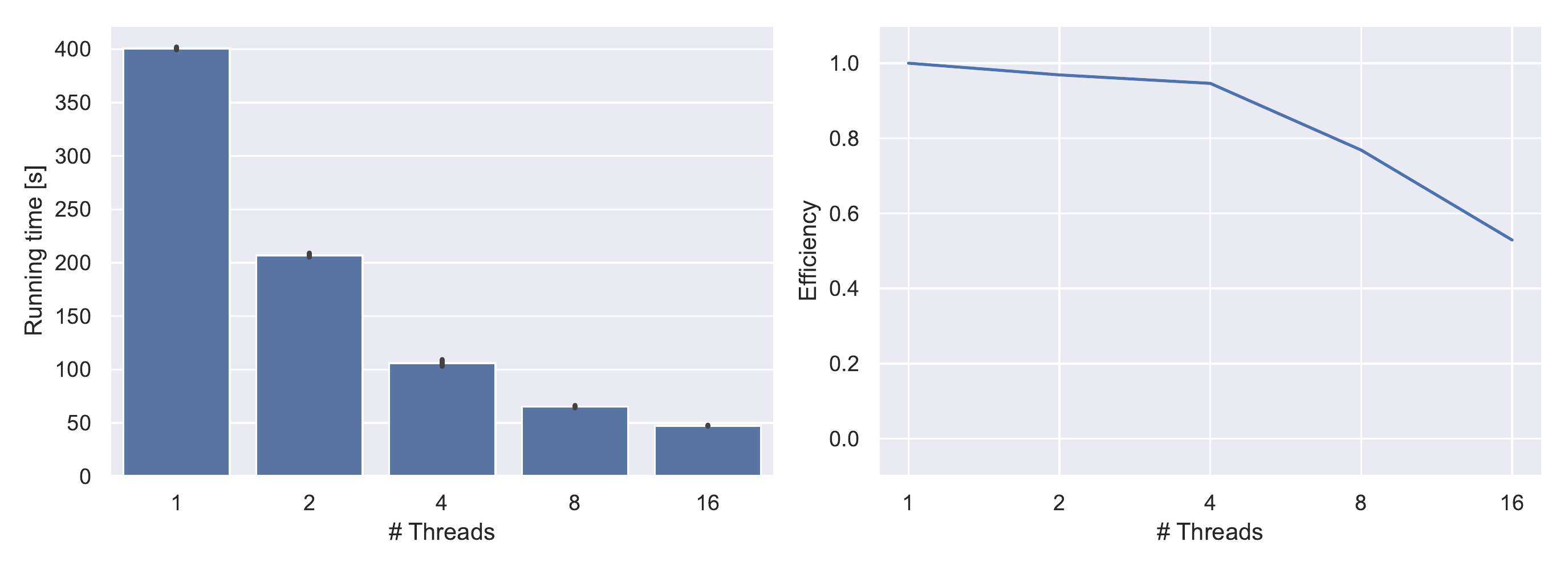}
\caption{
Left: Mean running times of compression of IMP buckets for PTV Europe from 96 to 16. The black lines (barely visible) indicates the standard deviation.
Right: Parallel efficiency, i.e.\ speedup over single threaded running time divided by number of threads.
}\label{fig:comp_par}
\end{figure}

Figure~\ref{fig:comp_par} depicts running times and parallel efficiency of the compression algorithm with a varying number of threads on PTV Europe when merging 96 buckets until only 16 buckets remain.
Without parallelization, the compression takes 6 to 7 minutes.
This is almost as long as the full MMP preprocessing.
Luckily, with 16 threads, the running time can be reduced to 50 seconds which is a speedup of about 8.
With fewer threads, the efficiency is even higher.
Even though the parallelization does not scale perfectly, running times are still reduced significantly.
As a result, compression times only make up a small fraction of the total preprocessing times.


\begin{thebibliography}{10}

\bibitem{bdgmpsww-rptn-16}
Hannah Bast, Daniel Delling, Andrew~V. Goldberg, Matthias
  {M{\"u}ller--Hannemann}, Thomas Pajor, Peter Sanders, Dorothea Wagner, and
  Renato~F. Werneck.
\newblock {Route Planning in Transportation Networks}.
\newblock In Lasse Kliemann and Peter Sanders, editors, {\em Algorithm
  Engineering - Selected Results and Surveys}, volume 9220 of {\em Lecture
  Notes in Computer Science}, pages 19--80. Springer, 2016.

\bibitem{bgsv-mtdtt-13}
Gernot~Veit Batz, Robert Geisberger, Peter Sanders, and Christian Vetter.
\newblock {Minimum Time-Dependent Travel Times with Contraction Hierarchies}.
\newblock {\em ACM Journal of Experimental Algorithmics}, 18(1.4):1--43, April
  2013.

\bibitem{bdpw-dtdrp-16}
Moritz Baum, Julian Dibbelt, Thomas Pajor, and Dorothea Wagner.
\newblock {Dynamic Time-Dependent Route Planning in Road Networks with User
  Preferences}.
\newblock In {\em Proceedings of the 15th International Symposium on
  Experimental Algorithms (SEA'16)}, volume 9685 of {\em Lecture Notes in
  Computer Science}, pages 33--49. Springer, 2016.

\bibitem{bsw-rttau-19}
Valentin Buchhold, Peter Sanders, and Dorothea Wagner.
\newblock {Real-time Traffic Assignment Using Engineered Customizable
  Contraction Hierarchies}.
\newblock {\em ACM Journal of Experimental Algorithmics}, 24(2):2.4:1--2.4:28,
  2019.
\newblock URL: \url{https://dl.acm.org/citation.cfm?id=3362693}.

\bibitem{dgpw-crprn-13}
Daniel Delling, Andrew~V. Goldberg, Thomas Pajor, and Renato~F. Werneck.
\newblock {Customizable Route Planning in Road Networks}.
\newblock {\em Transportation Science}, 51(2):566--591, 2017.
\newblock URL: \url{http://dx.doi.org/10.1287/trsc.2014.0579}.

\bibitem{dn-crdtd-12}
Daniel Delling and Giacomo Nannicini.
\newblock {Core Routing on Dynamic Time-Dependent Road Networks}.
\newblock {\em Informs Journal on Computing}, 24(2):187--201, 2012.

\bibitem{dsw-cch-15}
Julian Dibbelt, Ben Strasser, and Dorothea Wagner.
\newblock {Customizable Contraction Hierarchies}.
\newblock {\em ACM Journal of Experimental Algorithmics}, 21(1):1.5:1--1.5:49,
  April 2016.
\newblock URL: \url{http://doi.acm.org/10.1145/2886843}.

\bibitem{d-ntpcg-59}
Edsger~W. Dijkstra.
\newblock {A Note on Two Problems in Connexion with Graphs}.
\newblock {\em Numerische Mathematik}, 1(1):269--271, 1959.

\bibitem{gssv-erlrn-12}
Robert Geisberger, Peter Sanders, Dominik Schultes, and Christian Vetter.
\newblock {Exact Routing in Large Road Networks Using Contraction Hierarchies}.
\newblock {\em Transportation Science}, 46(3):388--404, August 2012.

\bibitem{gh-cspas-05}
Andrew~V. Goldberg and Chris Harrelson.
\newblock {Computing the Shortest Path: {A*} Search Meets Graph Theory}.
\newblock In {\em Proceedings of the 16th Annual {ACM--SIAM} Symposium on
  Discrete Algorithms (SODA'05)}, pages 156--165. SIAM, 2005.

\bibitem{gw-cppsp-05}
Andrew~V. Goldberg and Renato~F. Werneck.
\newblock {Computing Point-to-Point Shortest Paths from External Memory}.
\newblock In {\em Proceedings of the 7th Workshop on Algorithm Engineering and
  Experiments (ALENEX'05)}, pages 26--40. SIAM, 2005.

\bibitem{ghuw-fbndocch-19}
Lars Gottesb{\"u}ren, Michael Hamann, Tim~Niklas Uhl, and Dorothea Wagner.
\newblock {Faster and Better Nested Dissection Orders for Customizable
  Contraction Hierarchies}.
\newblock {\em Algorithms}, 12(9):196, 2019.
\newblock URL: \url{https://doi.org/10.3390/a12090196}.

\bibitem{hnr-afbhd-68}
Peter~E. Hart, Nils Nilsson, and Bertram Raphael.
\newblock {A Formal Basis for the Heuristic Determination of Minimum Cost
  Paths}.
\newblock {\em IEEE Transactions on Systems Science and Cybernetics},
  4:100--107, 1968.

\bibitem{bingblog}
Bing maps new routing engine.
\newblock
  \url{https://blogs.bing.com/maps/2012/01/05/bing-maps-new-routing-engine/}.
\newblock Accessed: 2020-01-25.

\bibitem{or-tnp-89}
Ariel Orda and Raphael Rom.
\newblock {Traveling without waiting in time-dependent networks is NP-hard}.
\newblock Technical report, Dept. Electrical Engineering, Technion-Israel
  Institute of Technology, 1989.

\bibitem{ss-hhhes-05}
Peter Sanders and Dominik Schultes.
\newblock {Highway Hierarchies Hasten Exact Shortest Path Queries}.
\newblock In {\em Proceedings of the 13th Annual European Symposium on
  Algorithms (ESA'05)}, volume 3669 of {\em Lecture Notes in Computer Science},
  pages 568--579. Springer, 2005.

\bibitem{swz-umlgt-02}
Frank Schulz, Dorothea Wagner, and Christos Zaroliagis.
\newblock {Using Multi-Level Graphs for Timetable Information in Railway
  Systems}.
\newblock In {\em Proceedings of the 4th Workshop on Algorithm Engineering and
  Experiments (ALENEX'02)}, volume 2409 of {\em Lecture Notes in Computer
  Science}, pages 43--59. Springer, 2002.

\bibitem{strasser:OASIcs:2017:7897}
Ben Strasser.
\newblock {Dynamic Time-Dependent Routing in Road Networks Through Sampling}.
\newblock In Gianlorenzo D'Angelo and Twan Dollevoet, editors, {\em 17th
  Workshop on Algorithmic Approaches for Transportation Modelling,
  Optimization, and Systems (ATMOS 2017)}, volume~59 of {\em OpenAccess Series
  in Informatics (OASIcs)}, pages 3:1--3:17, Dagstuhl, Germany, 2017. Schloss
  Dagstuhl--Leibniz-Zentrum fuer Informatik.
\newblock URL: \url{http://drops.dagstuhl.de/opus/volltexte/2017/7897}, \href
  {https://doi.org/10.4230/OASIcs.ATMOS.2017.3}
  {\path{doi:10.4230/OASIcs.ATMOS.2017.3}}.

\bibitem{swz-sfert-21}
Ben Strasser, Dorothea Wagner, and Tim Zeitz.
\newblock {Space-efficient, Fast and Exact Routing in Time-Dependent Road
  Networks}.
\newblock {\em Algorithms}, 14(3), January 2021.
\newblock URL: \url{https://www.mdpi.com/1999-4893/14/3/90}.

\bibitem{strasser_et_al:LIPIcs.SEA.2021.6}
Ben Strasser and Tim Zeitz.
\newblock {A Fast and Tight Heuristic for A* in Road Networks}.
\newblock In David Coudert and Emanuele Natale, editors, {\em 19th
  International Symposium on Experimental Algorithms (SEA 2021)}, volume 190 of
  {\em Leibniz International Proceedings in Informatics (LIPIcs)}, pages
  6:1--6:16, Dagstuhl, Germany, 2021. Schloss Dagstuhl -- Leibniz-Zentrum
  f{\"u}r Informatik.
\newblock URL: \url{https://drops.dagstuhl.de/opus/volltexte/2021/13778}, \href
  {https://doi.org/10.4230/LIPIcs.SEA.2021.6}
  {\path{doi:10.4230/LIPIcs.SEA.2021.6}}.

\bibitem{wz-cplttdap-22}
Nils Werner and Tim Zeitz.
\newblock {Combining Predicted and Live Traffic with Time-Dependent A*
  Potentials}.
\newblock In {\em Proceedings of the 30th Annual European Symposium on
  Algorithms (ESA'22)}, Leibniz International Proceedings in Informatics,
  September 2022.

\bibitem{z-nphsp-22}
Tim Zeitz.
\newblock {NP-Hardness of Shortest Path Problems in Networks with Non-FIFO
  Time-Dependent Travel Times }.
\newblock {\em Information Processing Letters}, May 2022.
\newblock URL: \url{https://doi.org/10.1016/j.ipl.2022.106287}.

\end{thebibliography}
\end{document}